\documentstyle[12pt]{article}
\begin{document}

\title{Doubly Lopsided Mass Matrices from Supersymmetric SU(N) Unification}

\author{{\bf S.M. Barr} \\
Bartol Research Institute \\ University of Delaware \\
Newark, Delaware 19716}

\date{\today}

\maketitle

\begin{abstract}
It is shown that in supersymmetric $SU(N)$ models with $N>5$ the so-called ``doubly lopsided" mass matrix structure can emerge in a natural way.  The non-trivial flavor structure is entirely accounted for by the $SU(N)$ gauge symmetry and supersymmetry,
without any ``flavor symmetry". The hierarchy among the families results
directly from a hierarchy of scales in the chain of breaking from
$SU(N)$ to the Standard Model group. A simple $SU(7)$ example is presented.
\end{abstract}

\newpage

\section{Introduction}

In a recent paper, it was shown that grand unified theories (GUTs) based on the groups
$SU(N)$, with $N>5$, can lead to a non-trivial flavor structure for the known quarks and leptons even in the absence of flavor symmetries \cite{su8}. The central point is that the different
families, which all transform in the same way under either the standard model group
$G_{SM} \equiv SU(3)_c \times SU(2)_L \times U(1)_Y$ or
$SU(5)$, can transform differently under $SU(N)$
\cite{sun}. Since they transform differently, the mixing of the families is inhibited
by $SU(N)$. Thus it is the ``vertical" unification group 
rather than a ``horizontal" symmetry that distinguishes the families and produces
a non-trivial flavor structure. Moreover, the flavor
structure that typically arises for the most economical sets of quarks and leptons is
a ``doubly lopsided" one
\cite{doublylop}, which is known to reproduce many of the qualitative
features of the observed pattern of quark and lepton masses and mixings.

In Ref. 1, a non-supersymmetric $SU(8)$ model
was presented that illustrated these ideas.  An appealing feature of
non-supersymmetric models of this type is that the family hierarchy
can be a ``radiative" one, with the heaviest
quarks and leptons (maybe only the top quark) getting mass from
renormalizable tree-level terms and
the lighter quarks and leptons getting mass from higher-dimension operators
induced by loops. However, if one supersymmetrizes such models, these loops
are suppressed by factors of $M_{SUSY}/M_{GUT}$ due to the non-renormalization
theorems. In this paper we look at supersymmetric $SU(N)$ models in which
there is a non-radiative fermion mass hierarchy. In these models the smaller
elements of the quark and lepton mass matrices are suppressed by powers of
$M_N/M_{P \ell}$, where $M_N$ is a symmetry-breaking scale associated with
the breaking of $SU(N)$ down to $G_{SM}$. There can be several such scales.
In fact, in the $SU(7)$ example that will be described, one
gauge-symmetry-breaking scale controls the masses of
the second family and another controls the masses of the first family.
An interesting feature, then, of the kinds of models we are proposing is that
a hierarchy in the scales of breaking of the ``vertical" unification group is directly reflected
in the ``horizontal" mass hierarchy among the families.

The paper is organized as follows. In section 2, a very brief review will be
given of the basic idea of doubly lopsided models and of how $SU(N)$ GUTs naturally
produce a hierarchical and doubly lopsided structure.  In section 3,
a simple supersymmetric $SU(7)$ model is described and the roles played
by the grand unified symmetry and supersymmetry in restricting the form of the mass
matrices is explained.
In section 4, a different version of the $SU(7)$ model is briefly discussed, in which only
the top quark obtains mass from renormalizable Yukawa terms and the other fermions get
mass from higher-dimension, Planck-scale-suppressed operators.
The final section gives a summary and conclusions.

\section{Lopsided models and $SU(N)$ unification}

The basic idea of doubly lopsided models \cite{doublylop} is that there is large
mixing among the three $\overline{{\bf 5}}$ multiplets of quarks and
leptons and small mixing among the three ${\bf 10}$ multiplets.  (We use
$SU(5)$ language in discussing the fermion multiplets for
convenience. The chain of breaking of the grand unified group need
not actually go through $SU(5)$.) For example, suppose the mixing between the
${\bf 10}_2$ and ${\bf 10}_3$ is of order $\epsilon \ll 1$, the
mixing between ${\bf 10}_1$ and ${\bf 10}_2$ is of order $\delta \ll
1$, and the mixings among the $\overline{{\bf 5}}$'s are all of order 1.
Then, since the mass matrices of the up-type quarks, down-type
quarks, charged leptons, and light neutrinos appear in terms
respectively of the form, ${\bf 10}_i (M_U)_{ij} {\bf 10}_j$, ${\bf 10}_i
(M_D)_{ij} \overline{{\bf 5}}_j$, $\overline{{\bf 5}}_i (M_L)_{ij}
{\bf 10}_j$, and $\overline{{\bf 5}}_i (M_{\nu})_{ij} \overline{{\bf
5}}_j$, these matrices have the form

\begin{equation}
\begin{array}{ll}
M_U \sim \left( \begin{array}{ccc} \delta^2 \epsilon^2 & \delta \epsilon^2 &
\delta \epsilon \\ \delta \epsilon^2 & \epsilon^2 & \epsilon \\
\delta \epsilon & \epsilon & 1 \end{array} \right) m, \;\;\; &
M_D \sim \left( \begin{array}{ccc} \delta \epsilon & \delta \epsilon &
\delta \epsilon \\ \epsilon & \epsilon & \epsilon \\
1 & 1 & 1 \end{array} \right) m', \\ & \\
M_{\nu} \sim \left( \begin{array}{ccc} 1 & 1 & 1 \\
1 & 1 & 1 \\
1 & 1 & 1 \end{array} \right) m_{\nu}, \;\;\; &
M_L \sim \left( \begin{array}{ccc} \delta \epsilon & \epsilon & 1 \\
\delta \epsilon & \epsilon & 1 \\
\delta \epsilon & \epsilon & 1 \end{array} \right) m'.
\end{array}
\end{equation}

\noindent
Note that the mass matrices in
Eq. (1) are written in the convention that they are multiplied from the left
by the left-handed fermions and from the right by right-handed fermions.
The $\sim$ symbol means that only the order in $\delta$ and $\epsilon$ of the matrix
elements is given.
Inspection of these matrices shows that there is a relatively steep
mass hierarchy among the up-type-quark masses ($m_u/m_c \sim
\delta^2$, $m_c/m_t \sim \epsilon^2$), a less steep hierarchy among
the down-type-quark masses and among the charged-lepton masses
($m_d/m_s \sim m_e/m_{\mu} \sim \delta$, $m_s/m_b \sim
m_{\mu}/m_{\tau} \sim \epsilon$), and a very weak hierarchy among
the neutrino masses. These predictions correspond to what is
observed. Inspection of the mass matrices also reveals that there are
small mixing
angles for the left-handed quarks, since they are in the ${\bf 10}$'s
($V_{cb} \sim \epsilon$, $V_{us}
\sim \delta$, $V_{ub} \sim \delta \epsilon$) and large ($O(1)$)
mixing angles for the left-handed leptons, since they are in the $\overline{{\bf 5}}$'s.
This pattern of angles also corresponds to
what is observed.

As Eq. (1) shows, the mass matrices of the down-type quarks and
charged leptons are highly asymmetrical,
which is the reason for the name ``lopsided".  In (singly) lopsided models
\cite{singlylop}, \cite{lopFV} only the
23 and 32 elements of these mass matrices are highly asymmetrical, explaining the fact that
the large atmospheric neutrino mixing angle is large while the corresponding quark
mixing angle $V_{cb}$ is small. ($U_{\mu 3} \equiv \sin \theta_{atm} \sim 1$,
$V_{cb} \sim \epsilon$.) In the doubly lopsided models, there is
assumed also to exist
a large asymmetry in the 13 and 31 elements, as in Eq. (1).  This asymmetry explains the
fact that the solar neutrino angle is large while the corresponding quark angle
$\sin \theta_C = V_{us}$ is small. ($U_{e2} \equiv \sin \theta_{sol} \sim 1$,
$V_{us} \sim \delta$.)

As explained in Ref. 1, the doubly lopsided pattern emerges naturally
in $SU(N)$ grand unification, even without any flavor symmetry. The reason has to do with the
embedding of the three families in the multiplets of $SU(N)$. If the
three families of light quarks and leptons come from antisymmetric tensor
representations of $SU(N)$, then the
most economical way to cancel $SU(N)$ anomalies with a
specified number of families in the low-energy spectrum
is to have a few larger
tensors (i.e. rank $> 1$) plus many anti-fundamental representations.
Some examples will be given later.
The ${\bf 10}$'s of $SU(5)$ are contained in the larger tensors and typically
transform differently under the $SU(N)$, whereas typically all the
$\overline{{\bf 5}}$'s of $SU(5)$ are contained in the
anti-fundamentals of $SU(N)$ and therefore transform in exactly the
same way under $SU(N)$. Since the $\overline{{\bf 5}}$'s are not
distinguished from each other by the symmetries of the theory, they
tend naturally to mix strongly with each other, whereas the
${\bf 10}$'s can only mix with each other to the extent that the symmetries
of $SU(N)$ that distinguish them are broken, and therefore their mixing is suppressed.

In the model of Ref. 1, for example, the quarks and leptons are
contained in the anomaly free set of $SU(8)$ representations ${\bf
56} + {\bf 28} + 9 (\overline{{\bf 8}}) = \psi^{[ABC]} + \psi^{[AB]}
+ \psi_{(m)A}$, where the indices $A,B,C$ ($= 1,..., 8$) are $SU(8)$
indices, while $m$ ($= 1,...,9$) just labels the nine
anti-fundamental representations. In $SU(8)$, this is the most
economical anomaly-free set of fermions that gives three families.
When this set is decomposed under $SU(5)$ it gives $4 ({\bf 10}) +
\overline{{\bf 10}} + 9 (\overline{{\bf 5}}) + 6 ({\bf 5}) + 31
({\bf 1})$, which leaves a low-energy residue of three $({\bf 10} +
\overline{{\bf 5}})$ families after vectorlike pairs get superlarge
mass. The ${\bf 10}$'s are $\psi^{\alpha \beta}$, $\psi^{\alpha
\beta 6}$, $\psi^{\alpha \beta 7}$, and $\psi^{\alpha \beta 8}$
(where $\alpha, \beta = 1, ..., 5$ are $SU(5)$ indices), which
obviously transform differently under $SU(8)$. On the other hand,
the $\overline{{\bf 5}}$'s are all of the form $\psi_{(m) \alpha}$
and are not distinguished in any way by the gauge symmetries.

\section{An illustrative supersymmetric model}

The model we will study in this paper has gauge group $SU(7)$ and quarks and
leptons in the following multiplets: ${\bf 35} + 2 ({\bf 21}) +
8 (\overline{{\bf 7}}) = \psi^{[ABC]} + \psi^{[AB]}_{(a)} +
\psi_{(m)A}$, where $A,B,C$ ($= 1,...,7$) are $SU(7)$ indices, and $a$ ($=
1,2$) and $m$ ($=1,...,8$) are labels distinguishing multiplets of the
same type. When decomposed under $SU(5)$ (which it is convenient to
use to classify the fermions, even if the chain of symmetry breaking
does not go through $SU(5)$), this gives

\begin{equation}
\begin{array}{ccl}
\psi^{[ABC]} = {\bf 35} & \rightarrow & \overline{{\bf 10}} + 2 ({\bf 10}) + {\bf 5} \\
& = & \psi^{\alpha \beta \gamma} + \psi^{\alpha \beta I} + \psi^{\alpha 67}, \\ & & \\
\psi^{[AB]}_{(a)} = 2 ({\bf 21}) & \rightarrow & 2 ({\bf 10}) +
4 ({\bf 5}) + 2 ({\bf 1}) \\
& = & \psi^{\alpha \beta}_{(a)} + \psi^{\alpha I}_{(a)} + \psi^{67}_{(a)}, \\ & & \\
\psi_{(m)A} = 8 (\overline{{\bf 7}}) & \rightarrow & 8 (\overline{{\bf 5}}) +
16 ({\bf 1}) \\
& = & \psi_{(m) \alpha} + \psi_{(m) I},
\end{array}
\end{equation}

\noindent
where $\alpha, \beta, \gamma = 1,...,5$ are $SU(5)$ indices and $I = 6,7$ are
$SU(2)'$ indices of the $SU(5) \times SU(2)' \times U(1)'$ subgroup of $SU(7)$.
(This is one of the most economical three-family sets of $SU(7)$ fermions, having
a total of 11 multiplets with 133 components. Another economical set, which gives 
a similar model, is $2({\bf 35}) + {\bf 21} + 7(\overline{{\bf 7}})$,
which has 10 multiplets and 140 components. The set with the fewest components 
is $3({\bf 21}) + 9(\overline{{\bf 7}})$, which has 12 multiplets with 126
components. The set with the fewest multiplets is $3({\bf 35}) + 6(\overline{{\bf 7}})$, which has 9 multiplets with 147 components. However, in these last two possibilities 
there is simply a triplication of multiplets, so that $SU(7)$ does not distinguish
among the families. These four sets are the most economical by far, the next simplest sets
having 189 and 196 components. So of the four simplest possibilities in $SU(7)$, two 
give models of the type being proposed.)

The Higgs content of the model consists of the following types of Higgs
superfields (there can be several of each type): adjoint multiplets $\Omega^A_B$, plus the totally antisymmetric tensor
multiplets $H^A$, $\overline{H}_A$, $H^{[AB]}$,
$\overline{H}_{[AB]}$, $H^{[ABC]}$, and
$\overline{H}_{[ABC]}$.  All the $G_{SM}$-singlet
components of these Higgs multiplets are assumed to have superlarge VEVs, namely
$H^I$ and $\overline{H}_I$ ($I=6,7$),
$H^{67}$, $\overline{H}_{67}$, and $\Omega^A_A$. All the components that transform
under $G_{SM}$ in the same way as the neutral components of $H_u$ and $H_d$ of the
MSSM are
assumed to get weak-scale VEVs, namely $H^2$, $H^{2I}$ ($I = 6,7$), $H^{267}$,
$\overline{H}_2$, $\overline{H}_{2I}$ ($I = 6,7$), and $\overline{H}_{267}$.
(We use the convention that $\alpha = 1,2$ are $SU(2)_L$ indices, and $\alpha = 3,4,5$
are $SU(3)_c$
indices.) It should be noted that there must be a ``matter parity" symmetry to
distinguish ``Higgs" multiplets from ``matter" (i.e. quark and lepton) multiplets.
This is typical of supersymmetric models. However, no ``flavor" symmetry exists
that distinguishes among the matter multiplets or among the Higgs multiplets.

The most general renormalizable Yukawa superpotential contains the following
couplings
(where we use the obvious notation that $[p]$ refers to a rank-$p$ antisymmetric tensor
and $\overline{[p]}$ refers to its conjugate, the subscript $L$ refers to a left-handed
supermultiplet
of quarks and leptons, and the subscript $H$ refers to a Higgs supermultiplet):
the

\begin{equation}
\begin{array}{ccl}
([3]_L [2]_L) [2]_H & = & a_a (\psi^{ABC} \psi^{DE}_{(a)}) H^{FG} \epsilon_{ABCDEFG}, \\
& & \\
([3]_L \overline{[1]}_L) \overline{[2]}_H & = & b_m (\psi^{ABC} \psi_{(m)A})
\overline{H}_{BC}, \\
& & \\
([2]_L [2]_L) [3]_H & = & c_{ab} (\psi^{AB}_{(a)} \psi^{CD}_{(b)}) H^{EFG} \epsilon_{ABCDEFG}, \\
& & \\
([2]_L \overline{[1]}_L) \overline{[1]}_H & = & d_{am} (\psi^{AB}_{(a)} \psi_{(m)A})
\overline{H}_B, \\
& & \\
(\overline{[1]}_L \overline{[1]}_L) [2]_H & = & e_{mn} (\psi_{(m)A} \psi_{(n)B}) H^{AB}.
\\ & &
\end{array}
\end{equation}

Even though it is assumed that there may several copies
of Higgs multiplets of the same type, no index has been used to distinguished
among them in Eq. (3).
Note that there is no $([3]_L [3]_L) [1]_H$ term listed in Eq. (3), since
such a term vanishes identically by the antisymmetry of $[3]_L$. 

Of the terms in Eq. (3), only the first contributes to a superlarge mass term for
the $SU(5)$ $\overline{{\bf 10}}$ of fermions that sits in the $[3]_L$ multiplet.
In particular, this term contains $a_a (\psi^{\alpha \beta \gamma}
\psi^{\delta \epsilon}_{(a)}) H^{67} \epsilon_{\alpha \beta \gamma \delta \epsilon 67}$, which ``mates" the $\overline{{\bf 10}} = \psi^{\alpha \beta \gamma}$ to one linear
combination of
the two $\psi_{(a)}^{\delta \epsilon}$, leaving the other linear combination light.
Without loss of generality, one can define the superheavy linear combination to be
$\psi_{(2)}^{\delta \epsilon}$.
Altogether, there remain three light ${\bf 10}$'s,
which it will be convenient to denote by
${\bf 10}_1 \equiv \psi^{\delta \epsilon 7}$, ${\bf 10}_2 \equiv
\psi^{\delta \epsilon 6}$, and ${\bf 10}_3 \equiv
\psi_{(1)}^{\delta \epsilon}$.

Turning now to the superlarge masses of the $\overline{{\bf 5}}$'s and ${\bf 5}$'s,
one sees that the term $([3]_L \overline{[1]}_L) \overline{[2]}_H$ in Eq. (3)
contains $b_m (\psi^{\alpha 6 7} \psi_{(m) \alpha})
\overline{H}_{67}$, which couples the ${\bf 5}$ in the rank-3 tensor to a
$\overline{{\bf 5}}$, and the term
$([2]_L \overline{[1]}_L) \overline{[1]}_H$ contains $d_{am} (\psi^{\alpha I}_{(a)} \psi_{(m) \alpha})
\overline{H}_I$, which couples the ${\bf 5}$'s in the rank-2 tensor to
$\overline{{\bf 5}}$'s. (If there is only one $\overline{[1]}_H$ multiplet, then
not all of the ${\bf 5}$'s in the rank-2 tensor get mass from this term; however, we will assume in the model presented this section that there are at least two
$\overline{[1]}_H$ multiplets.) What remains light are
three $\overline{{\bf 5}}$'s, all of which come, of course, from the anti-fundamentals of $SU(N)$ and have the form
$\psi_{(m) \alpha}$. Without loss of generality one can define
the three light $\overline{{\bf 5}}$'s
to be $\overline{{\bf 5}}_1 \equiv \psi_{(1) \alpha}$,
$\overline{{\bf 5}}_2 \equiv \psi_{(2) \alpha}$, and
$\overline{{\bf 5}}_3 \equiv \psi_{(3) \alpha}$.

One is now in a position to understand how the weak-scale masses of the light
quarks and leptons arise in this model. First consider the
${\bf 10}$ to ${\bf 10}$ couplings that give mass to the up-type quarks.
The third term in Eq. (3), namely  $([2]_L [2]_L) [3]_H$, contains
$c_{11} (\psi^{\alpha \beta}_{(1)} \psi^{\gamma \delta}_{(1)})
H^{267} \epsilon_{\alpha \beta \gamma \delta 2 6 7}$. Since
$\psi^{\alpha \beta}_{(1)} \equiv {\bf 10}_3$, this term
gives a 33 element to $M_U$, the mass matrix of the up-type
quarks.

The first term in Eq. (3), namely $([3]_L [2]_L) [2]_H$, contains
$a_1 [(\psi^{\alpha \beta 6} \psi^{\gamma \delta}_{(1)}) H^{27}
- (\psi^{\alpha \beta 7} \psi^{\gamma \delta}_{(1)}) H^{26}]
\epsilon_{\alpha \beta \gamma \delta 2 6 7}$, which give contributions to
the 13, 31, 23, and 32 elements of $M_U$. (It should be noted that if there
were only one $H^{AB}$ multiplet, then $([3]_L[2]_L)[2]_H$
would only involve
the superheavy field $\psi^{\gamma \delta}_{(2)}$, not the light field
$\psi^{\gamma \delta}_{(1)}$. This is not so, however, if more than one
$H^{AB}$ multiplet exists, as is assumed in the model described in this section.)

There is no renormalizable Yukawa term that couples
${\bf 10}_1 \equiv \psi^{\delta \epsilon 7}$ and ${\bf 10}_2 \equiv
\psi^{\delta \epsilon 6}$ to themselves or to each other. The only
renormalizable term that could do so would be
a $([3]_L [3]_L) [1]_H$ coupling
$(\psi^{\alpha \beta 6} \psi^{\gamma \delta 7}) H^2
\epsilon_{\alpha \beta \gamma \delta 2 6 7}$; however, as already noted, this vanishes identically by the
antisymmetry of the indices. Thus, the most general
set of renormalizable terms consistent with $SU(7)$ and supersymmetry gives
a $M_U$ of the form

\begin{equation}
M_U \sim \left( \begin{array}{ccc}
0 & 0 & \delta \epsilon \\ 0 & 0 & \epsilon \\
\delta \epsilon & \epsilon & 1 \end{array} \right) v_u,
\end{equation}

\noindent
This matrix has been written in a way that suggests that the 13 and 31 elements
of $M_U$ are
much smaller than the 23 and 32 elements, and that they in turn are much smaller than
the 33 element, as in Eq. (1).  This would not generally
be the case, of course, but would be if
there were the following hierarchy
among the VEVs of the Higgs fields:

\begin{equation}
\begin{array}{rcc}
\langle H^2 \rangle, \langle H^{267} \rangle & \sim & v_u \\
\gg \langle H^{27} \rangle & \sim & \epsilon v_u \\
\gg \langle H^{26} \rangle & \sim & \delta \epsilon v_u,
\end{array}
\end{equation}

\noindent
It will be assumed that this hierarchy holds, as well as the related hierarchies

\begin{equation}
\begin{array}{rcc}
\langle \overline{H}_2 \rangle, \langle \overline{H}_{267} \rangle & \sim &
v_d \\
\gg \langle \overline{H}_{26} \rangle & \sim & \epsilon v_d \\
\gg \langle \overline{H}_{27} \rangle & \sim &
\delta \epsilon v_d,
\end{array}
\end{equation}

\noindent
and

\begin{equation}
\begin{array}{rcc}
\langle H^{67} \rangle, \langle \overline{H}_{67} \rangle & \sim & M_{P \ell} \\
\gg \langle H^7 \rangle, \langle \overline{H}_6 \rangle  & \sim & \epsilon M_{P \ell} \\
\gg \langle H^6 \rangle, \langle \overline{H}_7 \rangle  & \sim & \delta \epsilon
M_{P \ell}.
\end{array}
\end{equation}

\noindent
Such a hierarchy can be
understood in a group-theoretical way. The group $SU(7)$ contains the subgroup
$SU(5) \times SU(2)' \times U(1)'$, where $SU(5)$ contains the Standard Model
group $G_{SM}$ and $SU(2)'$ acts on the indices 6 and 7.  Denote the diagonal
generator of $SU(2)'$ (namely diag$(0,0,0,0,0,\frac{1}{2}, -\frac{1}{2})$)
by $I'_3$. Then the hierarchies in Eqs. (5) - (7) can be succinctly stated by saying
that components of left-handed superfields that have $I'_3 = \frac{1}{2}$ are
suppressed by a factor $\delta \epsilon$, those with
$I'_3 = -\frac{1}{2}$ are suppressed by a factor $\epsilon$, and
those with $I'_3 = 0$ are unsuppressed. Later it will be sees that such a pattern
can naturally arise in the context of supersymmetry. (The $SU(2)'$ subgroup of 
$SU(7)$ is playing the role of a flavor group, under which the
${\bf 10}$'s of $SU(5)$ of the lightest two families transform as doublets.
It is interesting that many models have been proposed in which the three families transform as ${\bf 2} + {\bf 1}$ under a flavor symmetry that is either $SU(2)$ or a discrete
subgroup of $SU(2)$ \cite{su2family}.)

Note that the form of the matrix in Eq. (4) has rank = 2, implying that
the renormalizable Yukawa terms leave the $u$ quark massless. In fact, this is the
only quark or lepton (besides the light neutrinos) that must obtain mass from
higher-dimension operators. This corresponds to the fact that the ratio
$m_u/m_t$ is by far the smallest of all the interfamily mass ratios of the quarks and leptons.

The elements of the 12 block of $M_U$ (and thus $m_u$)
can be induced by higher-dimension operators such as
$(\psi^{ABC} \psi^{DEK}) (H^{FG} \overline{H}_K/M_{P \ell}) \epsilon_{ABCDEFG}$.
This operator gives, in
particular, the following terms:

\noindent
(a) $(\psi^{\alpha \beta 6} \psi^{\gamma \delta 6}) (H^{27} \overline{H}_6/M_{P \ell})
\epsilon_{\alpha \beta \gamma \delta 2 6 7}$, which
is a contribution to $(M_U)_{22}$ and is of order $\epsilon^2 v_u$;

\noindent
(b) $(\psi^{\alpha \beta 7} \psi^{\gamma \delta 7}) (H^{26} \overline{H}_7/M_{P \ell})
\epsilon_{\alpha \beta \gamma \delta 2 6 7}$, which
is a contribution to $(M_U)_{11}$ and is of order $\delta^2 \epsilon^2 v_u$; and

\noindent
(c) $(\psi^{\alpha \beta 6} \psi^{\gamma \delta 7}) ([H^{27} \overline{H}_7
- H^{26} \overline{H}_6]/M_{P \ell})
\epsilon_{\alpha \beta \gamma \delta 2 6 7}$, which
is a contribution to $(M_U)_{22}$ and is of order $\delta \epsilon^2 v_u$.
Thus, the matrix $M_U$ has the form

\begin{equation}
M_U \sim \left( \begin{array}{ccc} \delta^2 \epsilon^2 & \delta \epsilon^2 &
\delta \epsilon \\ \delta \epsilon^2 & \epsilon^2 & \epsilon \\
\delta \epsilon & \epsilon & 1 \end{array} \right) v_u,
\end{equation}

\noindent
as in Eq. (1).

The masses for the down-type quarks and charged leptons come from
${\bf 10}$ to $\overline{{\bf 5}}$ terms contained in the second and fourth terms of Eq. (3).
The term $([2]_L \overline{[1]}_L) \overline{[1]}_H$ contains
$d_{1m} (\psi^{\alpha 2}_{(1)} \psi_{(m)\alpha})
\overline{H}_2$.
This produces a mass coupling ${\bf 10}_3$ to
$\overline{{\bf 5}}_m$, $m=1,2,3$, and thus $3m$ elements of $M_D$, the
mass matrix of the down-type quarks, and $m3$ elements of $M_L$, the mass matrix of the charged leptons.
These elements are all of order $v_d$.

The term $([3]_L \overline{[1]}_L) \overline{[2]}_H$ contains
$b_m [(\psi^{\alpha 2 6} \psi_{(m) \alpha}) \overline{H}_{2 6} -
(\psi^{\alpha 2 7} \psi_{(m) \alpha}) \overline{H}_{2 7}]$
This gives mass terms coupling ${\bf 10}_2$ to
$\overline{{\bf 5}}_m$ that are of order $\epsilon v_d$ and mass terms
coupling ${\bf 10}_1$ to $\overline{{\bf 5}}_m$ that are of order $\delta \epsilon v_d$.

The resulting mass matrices have the form

\begin{equation}
M_D \sim \left( \begin{array}{ccc} \delta \epsilon & \delta \epsilon &
\delta \epsilon \\ \epsilon & \epsilon & \epsilon \\
1 & 1 & 1 \end{array} \right) v_d, \;\;\;
M_L \sim \left( \begin{array}{ccc} \delta \epsilon & \epsilon & 1 \\
\delta \epsilon & \epsilon & 1 \\
\delta \epsilon & \epsilon & 1 \end{array} \right) v_d.
\end{equation}

\noindent
as in Eq. (1). Actually, the operators considered above give the ``minimal $SU(5)$"
relations $M_D = M_L^T$. A breaking of this relation can result from higher-dimension terms
involving the adjoint Higgs fields. However, such terms, if induced by Planck-scale
effects, would be suppressed by $M_5/M_{P \ell}$, where $M_5$ is the scale at which
$SU(5)$ breaks. This is too small to give realistic ``Georgi-Jarlskog" factors
\cite{gj}.
However, such terms can easily be induced by integrating out fields that have mass
of order $M_5$ rather than $M_{P \ell}$. Another possibility is
that there are Higgs multiplets in the representation $H^A_{BCD}$, which
would allow renormalizable terms of
the form $(\psi^{\alpha 2 I} \psi_{(m) \alpha'}) H^{\alpha'}_{2I\alpha}$.
This is a generalization of having a ${\bf 45}_H$ contribute to fermion masses in
$SU(5)$ models, as in the original model of Georgi and Jarlskog \cite{gj}.

The question now arises whether the hierarchies among the VEVs shown
in Eqs. (5) - (7) are natural. First consider the superlarge VEVs
given in Eq. (7). Rather than doing a complete minimization with the
entire superpotential, it will suffice to consider the terms in the
superpotential that couple Higgs superfields of different ranks. For
example, consider the terms of the form $\alpha H^{AB}
\overline{H}_A \overline{H}_B + M H^A \overline{H}_A$. (If there
were only one anti-fundamental Higgs multiplet, then the term with
the coefficient $\alpha$ would vanish identically, because of the
antisymmetry of the indices $A$ and $B$. However, it is being
assumed that there are at least two such Higgs multiplets, so that
$\alpha$ and $M$ are really matrices. There should be an index on
$H_A$ to distinguish among the different copies, and indices on
$\alpha$ and $M$, but these indices have been suppressed.) We assume
that $\alpha$ is of order 1 and $M$ is of order $M_{P \ell}$.
Minimizing the superpotential with respect to $\overline{H}_A$ gives
the equation $H^A = - (M^{-1 T} \alpha) H^{AB} \overline{H}_B$.
Thus, if $\langle H^{67} \rangle \sim M_{P \ell}$, then $\langle H^6
\rangle \sim \langle \overline{H}_7 \rangle$ and $\langle H^7
\rangle \sim \langle \overline{H}_6 \rangle$, as in Eq. (7).

Note that supersymmetry plays a crucial role. The desired pattern of VEVs would
not be natural in a non-supersymmetric model, since in such a model the conjugate field $(\overline{H}^{\dag})^7$ could
be substituted anywhere for $H^7$, and thus minimization would tend to
give $\langle H^7 \rangle \sim \langle \overline{H}_7 \rangle$, and similarly
$\langle H^6 \rangle \sim \langle \overline{H}_6 \rangle$.

Turning to the $SU(2)_L$-doublets Higgs fields, one sees that there is a
four-by-four Higgs mass matrix ${\cal M}$

\begin{equation}
H {\cal M} \overline{H} = \left( H^i, H^{i67}, H^{i6}, H^{i7} \right)
\left( \begin{array}{cccc}
M_{P \ell} & \langle H^{67} \rangle & \langle H^6 \rangle & \langle H^7 \rangle \\
\langle \overline{H}_{67} \rangle & M_{P \ell} & \langle \overline{H}_7 \rangle &
\langle \overline{H}_6 \rangle \\
\langle \overline{H}_6 \rangle & \langle H^7 \rangle & M_{P \ell} &
\frac{\langle \overline{H}_6 \rangle \langle H^7 \rangle}{M_{P \ell}} \\
\langle \overline{H}_7 \rangle & \langle H^6 \rangle & \frac{\langle
\overline{H}_7 \rangle \langle H^6 \rangle}{M_{P \ell}} & M_{P \ell}
\end{array} \right) \left( \begin{array}{c} \overline{H}_i \\
\overline{H}_{i67} \\ \overline{H}_{i6} \\ \overline{H}_{i7}
\end{array} \right),
\end{equation}

\noindent
where $i$ is the $SU(2)_L$ index. In Eq. (10) we have not shown dimensionless
coefficients of order 1.
From the hierarchy in Eq. (7), it follows that

\begin{equation}
{\cal M} \sim \left( \begin{array}{cccc}
1 & 1 & \delta \epsilon & \epsilon \\
1 & 1 & \delta \epsilon & \epsilon \\
\epsilon & \epsilon & 1 & \epsilon^2 \\
\delta \epsilon & \delta \epsilon & \delta^2 \epsilon^2 & 1
\end{array} \right) M_{P \ell}.
\end{equation}

In this paper we do not address the question of a ``technically natural" solution of
the gauge hierarchy and doublet-triplet splitting problem. Rather, we simply assume that
${\cal M}$ is fine-tuned so that it has one weak-scale eigenvalue. (This could
be justified ``anthropically", for example in a landscape scenario \cite{anthropic}.)
First, consider the limit $\delta, \epsilon \rightarrow 0$. In that limit, one form
of ${\cal M}$ that has a weak-scale eigenvalue is

\begin{equation}
{\cal M} = \left( \begin{array}{cccc}
A A' & A B' & 0 & 0 \\
B A' & B B' & 0 & 0 \\
0 & 0 & C & 0 \\
0 & 0 & 0 & D \end{array}
\right) M_{P \ell} + O(M_{weak}),
\end{equation}

\noindent where $A, A', B, B', C, D \sim 1$. (${\cal M}$ could also
have a weak-scale eigenvalue if the 12 block did not have this
factorized form, but either $C$ or $D$ were of order the weak scale.
However, this possibility is not of interest for present purposes.)
If ${\cal M}$ has the form in Eq. (12), then the light Higgs
doublets are $(H_u)^i = (-B^* H^i + A^* H^{i67})/\sqrt{|A|^2 +
|B|^2}$ and $(H_d)_i = (-B^{\prime *} H_i + A^{\prime *} H_{i67})/
\sqrt{|A'|^2 + |B'|^2}$. Now, taking $\delta$, $\epsilon$ to be
non-zero but much less than 1, the form of Eq. (12) becomes

\begin{equation}
{\cal M} = \left( \begin{array}{cccc}
A A' + O( \delta \epsilon^2) & A B' + O( \delta \epsilon^2)& O( \delta \epsilon) &
O( \epsilon) \\
B A' + O( \delta \epsilon^2) & B B' + O( \delta \epsilon^2) & O( \delta \epsilon)&
O( \epsilon) \\
O( \epsilon) & O( \epsilon) & C + O( \delta \epsilon^2) & O( \epsilon^2) \\
O( \delta \epsilon) & O( \delta \epsilon) & O( \delta^2 \epsilon^2) & D
+ O( \delta \epsilon^2) \end{array}
\right) M_{P \ell} + O(M_{weak}),
\end{equation}

\noindent
from which it is easy to see that the light doublets are

\begin{equation}
\begin{array}{l}
(H_u)^i = \left( - B^* H^i + A^* H^{i67} + O(\delta \epsilon) H^{i6}
+ O(\epsilon) H^{i7} \right)/\sqrt{|A|^2 + |B|^2}, \\ \\
(H_d)_i = \left( - B^{\prime *} H_i + A^{\prime *} H_{i67} + O(\epsilon) H_{i6}
+ O(\delta \epsilon) H_{i7} \right)/\sqrt{|A'|^2 + |B'|^2}.
\end{array}
\end{equation}

\noindent If $\langle (H_u)^2 \rangle \equiv v_u$ and $\langle
(H_d)_2 \rangle \equiv v_d$, then

\begin{equation}
\begin{array}{l}
(\langle H^2 \rangle, \langle H^{267} \rangle, \langle
H^{26}\rangle, \langle H^{27} \rangle)
 = ( -\frac{B}{\sqrt{|A|^2 + |B|^2}}, \frac{A}{\sqrt{|A|^2 + |B|^2}}, O(\delta \epsilon),
O(\epsilon)) v_u, \\
(\langle H_2 \rangle, \langle H_{267} \rangle, \langle H_{26}
\rangle , \langle H_{27} \rangle)
 = ( -\frac{B'}{\sqrt{|A'|^2 + |B'|^2}}, \frac{A'}{\sqrt{|A'|^2 + |B'|^2}}, O(\epsilon),
O(\delta \epsilon)) v_d,
\end{array}
\end{equation}

\noindent
which is just the pattern assumed in Eqs. (5) and (6).

Since the model just outlined is supersymmetric and has a lopsided mass matrix
structure, the question of flavor violation arises. It is well-known that supersymmetric
lopsided models give larger flavor violation than non-lopsided models, due to
the large off-diagonal elements in the quark and lepton mass matrices \cite{lopFV}.
However, if supersymmetry is broken in a flavor-blind way, as in models
with gauge-mediated SUSY-breaking,
excessive flavor violation due to the lopsided is avoided.

\section{An alternative version of the model}

In the $SU(7)$ model described in the previous section, it was assumed that
there were several copies of certain Higgs multiplets. But it is
interesting to consider also the possibility that there is just one copy of each
antisymmetric tensor multiplet of Higgs fields. This has several consequences. First,
not all of the vectorlike pairs of $SU(5)$ multiplets would then get superheavy mass
from renormalizable Yukawa operators. In a non-supersymmetric model, they would
get superheavy mass from loops, as in the model described in Ref. 1.
In a model with low-energy supersymmetry such loops would be suppressed,
but those fermions can nonetheless get superheavy
mass from non-renormalizable operators induced by Planck-scale physics.
A second consequence of the more limited set of Higgs multiplets
is that most of the light quarks and leptons would not get weak-scale masses
from renormalizable Yukawa  operators.
Here again, operators induced by Planck-scale physics can generate these masses.

First, consider the masses of the superheavy ${\bf 10}$'s and
$\overline{{\bf 10}}$'s. The only renormalizable term that
contributes to these is the first term in Eq. (3). If there are
several $H^{AB}$ in the model--- denote them $H^{AB}_{(\lambda)}$ ---
as assumed in the previous section, then several linear combinations
of the $\psi_{(a)}^{AB}$ (namely $a^{\lambda}_a \psi^{AB}_{(a)}$)
appear in the first term in Eq. (3), so that in general both
$\psi^{AB}_{(1)}$ and $\psi^{AB}_{(2)}$ appear. However, if there is
only a single $H^{AB}$, then only one linear combination of
$\psi^{AB}_{(a)}$ appears in the first term of Eq. (3) (namely $a_a
\psi_{(a)}^{AB}$). Without loss of generality, one can call this
$\psi^{AB}_{(2)}$. Therefore, the first term of Eq. (3) contains
$a_2 \psi^{\alpha \beta \gamma} \psi_{(2)}^{\delta \epsilon} H^{67}
\epsilon_{\alpha \beta \gamma \delta \epsilon 67}$, which gives
$\psi^{AB}_{(2)}$ superheavy mass. If one assumes that this is the
only contribution to the superheavy masses of the ${\bf 10}$'s (i.e.
if one neglects contributions from higher-dimension operators), it
follows that $\psi^{\alpha \beta}_{(2)}$ is superheavy and
$\psi^{\alpha \beta}_{(1)}$ is light. As in the last section, denote the light
multiplet $\psi_{(1)}^{\alpha \beta}$ by ${\bf 10}_3$.
Since only $\psi_{(2)}^{AB}$ appears in the
first term of Eq. (3), that term makes no contribution to the masses
of the light ${\bf 10}$'s (if one neglects higher-dimension
operators).

In the model of the previous section, it was precisely the first term
in Eq. (3) that gave rise to the mass terms coupling ${\bf 10}_3$ to ${\bf 10}_1$ and
${\bf 10}_2$, terms such as $a_1 \psi^{\alpha \beta 6} \psi_{(1)}^{\gamma \delta} H^{2 7}
\epsilon_{\alpha \beta \gamma \delta 267}$.
Here, as just argued, this cannot happen, and so
the 13, 31, 23, and 32 elements of $M_U$ vanish if higher-dimension operators are
neglected. They do receive non-vanishing contributions, however, from various higher-dimension operators, such as
$(\psi^{\alpha \beta I} \psi^{\gamma \delta}_{(a)}) (H^{2K} \Omega^J_K/M_{P \ell}) \epsilon_{\alpha \beta \gamma \delta 2IJ}$ and
$(\psi^{\alpha \beta I} \psi^{\gamma \delta}_{(a)}) (H^{2JK} \overline{H}_K/M_{P \ell}) \epsilon_{\alpha \beta \gamma \delta 2IJ}$. The elements of the 12 block of $M_U$ also
arise from higher-dimension operators, though different ones.

By analogous reasoning, one can show that all of the elements of $M_D$ and $M_L$, the mass
matrices of the down-type quarks and charged leptons, vanish if
higher-dimension operators are neglected. The point is that the linear combinations of
the $\overline{{\bf 5}}$'s $\psi_{(m) \alpha}$ that appear in the terms of Eq. (3),
get superheavy masses, and so none of the light $\overline{{\bf 5}}$'s would
get weak-scale
masses from the terms in Eq. (3) alone.

One sees, then, an interesting feature of the version of the model we are
considering here, in which only a single copy of each Higgs multiplet exists:
only the top quark gets mass from a renormalizable term; all the other light fermions
get mass from higher-dimension operators.  This is what happens in the model of Ref. 1 also, except that there the higher-dimension operators came from loops and here they
come from Planck-scale physics. On the other hand, in the model of Ref. 1, since it is not supersymmetric, a hierarchy among the VEVs analogous to that given in
Eqs. (5)-(7) is not natural. Here, however, it can hold; and so, as in the version of
the model described in the previous section, the mass matrices have the form
given in Eqs. (8) and (9).

\section{Conclusions}

The group $SO(10)$ is often regarded as the most elegant for grand unification. Its
good features are that an entire family fits so neatly into one of its irreducible
multiplets and that it requires that right-handed neutrinos exist, unlike
$SU(5)$. However, if the families are simply placed in spinors of $SO(10)$, they
transform identically under the unification group, and one must introduce
flavor symmetries ad hoc in order to explain the non-trivial structure of the quark and lepton mass matrices. Virtually all published models of quark and lepton masses
have flavor symmetries in addition to the ``vertical" gauge group.

One of the beauties of $SU(N)$ unification, pointed out long ago \cite{sun},
is that the families do not in general transform in the same way under $SU(N)$
if $N > 5$. This creates the possibility, pointed out in \cite{su8}, of eliminating
flavor symmetries entirely. Nor does $SU(N)$ lack the supposed advantages of
$SO(10)$. $SU(N)$ unification gives an abundance of Standard Model-singlet fermions that play the role of right-handed neutrinos. Moreover, if it is assumed that
the quarks and leptons are in totally antisymmetric tensor multiplets of $SU(N)$, then
upon breaking to $SU(5)$ or $SU(3)_c \times SU(2)_L \times U(1)_Y$, one automatically
obtains a set of families just like those observed, with quantum numbers that
make them appear to come from $SO(10)$.

As noted in \cite{su8}, the smallest anomaly-free sets of fermions that give a fixed number of families typically have
a few larger tensors (rank $> 1$) and a many anti-fundamental multiplets. The
${\bf 10}$'s of $SU(5)$ are in the larger tensors and tend to transform
differently under $SU(N)$, which suppresses their mixing, whereas the
$\overline{{\bf 5}}$'s of $SU(5)$ are typically all in the anti-fundamentals and
transform in exactly the same way under $SU(N)$, which allows them to have large
mixings. This gives exactly the ``doubly lopsided" structure that is known to
reproduce well many of the features of the quark and lepton spectrum.

In this paper, a supersymmetric model has been constructed using the group
$SU(7)$. The quarks and leptons are in one of the most economical sets that gives three
families. No symmetry has been assumed except $SU(7)$, supersymmetry, and a
matter parity that distinguishes Higgs multiplets from matter (i.e. quark and lepton)
multiplets. No flavor symmetry is needed. An interesting feature of this model is that
the hierarchy among the families comes directly from a hierarchy of breaking scales of
the unified group. The hierarchy of $SU(7)$-breaking VEVs is of an interesting
type that requires supersymetry to achieve. $SU(7)$ breaks at near the Planck scale
to $SU(5) \times SU(2)'$. The $SU(2)'$ breaks in two stages, with VEVs of
components having
$I'_3 = +1/2$ being smaller than VEVs of components having $I'_3 = -1/2$. These
two ``small" scales (small compared to $M_{P \ell}$) control the mass scales of the first and second families.

It is rather remarkable that simple grand unified models, of a kind proposed very long
ago \cite{sun}, tend automatically to give a lopsided mass matrix structure
that yields small quark mixings, large neutrino mixings, and other features of the
quark and lepton spectrum that were unknown at that time.


\begin{thebibliography}{999}
\bibitem{su8} S.M. Barr, hep-ph/0804.1356
\bibitem{sun} H. Georgi, {\it Nucl. Phys.} {\bf B156}, 126 (1979);
S.M. Barr, {\it Phys.Rev.} {\bf D21}, 1424 (1980).
\bibitem{doublylop} K.S. Babu and S.M. Barr, {\it Phys. Lett.} {\bf B381}, 202 (1996);
N. Haba and H. Murayama, {\it Phys.Rev.} {\bf D63}, 053010 (2001);
K.S. Babu and S.M. Barr, {\it Phys.Lett.} {\bf B525}, 289 (2002);
S.M. Barr, ``Four Puzzles of Neutrino Mixing",
Talk given at 3rd Workshop on Neutrino Oscillations and Their Origin
(NOON 2001), Kashiwa, Japan, 5-8 Dec 2001, Published in {\it Kashiwa
2001, Neutrino oscillations and their origin} p. 358,
[hep-ph/0206085].
\bibitem{singlylop} C.H. Albright and S.M. Barr, {\it Phys.Rev.} {\bf D58}, 013002 (1998);
J. Sato and T. Yanagida, {\it Phys. Lett.} {\bf B430}, 127 (1998);
C.H. Albright, K.S. Babu, and S.M. Barr, {\it Phys.Rev.Lett.} {\bf 81}, 1167 (1998);
N. Irges, S. Lavignac, and P. Ramond, {\it Phys. Rev.} {\bf D58}, 035003 (1998);
J. Sato and T. Yanagida, {\it Phys.Lett.} {\bf B493}, 356 (2000);
T. Asaka, {\it Phys.Lett.} {\bf B562}, 291 (2003);
X.-D. Ji, Y.-C. Li, R.N. Mohapatra, {\it Phys.Lett.} {\bf B633},
755 (2006); 
\bibitem{lopFV} J. Sato and K. Tobe, {\it Phys.Rev.} D63:116010 (2001);
X.-J. Bi and Y.-B. Dai, {\it Phys.Rev.} D66:076006 (2002);
E. Jankowski and D.W. Maybury, {\it Phys.Rev.} D70:035004 (2004);
X.-D. Ji, Y. Li, and Y. Zhang, {\it Phys.Rev.} D75:055016 (2007).
\bibitem{su2family} R. Barbieri, G. Dvali, and L.J. Hall,
{\it Phys. Lett.} {\bf B377}, 76 (1996);
R. Barbieri, L.J. Hall, S. Raby, and A. Romanino,
{\it Nucl.Phys.} {B493}, 3 (1997).
\bibitem{gj} H. Georgi and C. Jarlskog, {\it Phys.Lett.} {\bf B86}, 297 (1979).
\bibitem{anthropic} V. Agrawal, S.M. Barr, J.F. Donoghue, and D. Seckel,
{\it Phys.Rev.} {\bf D57}, 5480 (1998); {\it Phys.Rev.Lett.} {\bf 80},
1822 (1998); S.M. Barr and A. Khan, {\it Phys.Rev.} D76:045002 (2007).
\end{thebibliography}
\end{document}